\documentclass[a4paper,11pt]{article}
\usepackage{latexsym,amssymb}
\begin{document}

\begin{center}
{\Large {\bf Exit probability and first passage time of a lazy Pearson walker:
Scaling behaviour}}\end{center}

\vskip 1cm

\begin{center}{\it Muktish Acharyya}\\
{\it Department of Physics, Presidency University,}\\
{\it 86/1 College Street, Calcutta-700073, INDIA}\\
{E-mail:muktish.physics@presiuniv.ac.in}\end{center}

\vskip 1cm

\noindent {\bf Abstract:} 
The motion of a lazy Pearson walker is studied
with different probability ($p$) of jump in two and three dimensions.
The probability of exit 
($P_e$) from a zone of
radius $r_e$, is studied as a function of $r_e$ with different values of
jump probability $p$. The exit probability $P_e$ is found to scale as
${P_e}p^{\alpha}=F({r_e}p^{\beta})$, 
which is obtained by 
method of data collapse. The first passage time ($t_1$) i.e., the time 
required for first exit from a zone is studied. The probability distribution
($P(t_1)$) of first passage time was studied for different values of jump
probability ($p$). The probability distribution of first passage time 
was found to scale as 
${P(t_1)}p^{\gamma} = G({t_1}p^{\delta})$. 
Where,
$F$ and $G$ are two scaling functions and $\alpha$, $\beta$, $\gamma$ and 
$\delta$ are some exponents. In both the dimensions, it is found that,
$\alpha = 0$, $\beta=-1/2$, $\gamma=-1$ and $\delta=1$.

\vskip 2cm


\newpage

\noindent {\bf I. Introduction:}

The random walk is a problem, studied widely in mathematics, statistics and physics
to analyse various natural phenomena. As an example,
in statistical physics, process of 
polymerization\cite{rev,grassberger}, diffusion\cite{jkb} 
of microparticles etc are some classic
phenomena, which have drawn much attention of the researcher in last
few decades. The basic mechanism of such phenomena are explained by
random walk\cite{thesis} in various form. Different kinds of random walks
 are studied on the
lattice in different dimensions by computer simulation. A few of them
may be mentioned here.
The absorbing phase transition in a conserved lattice gas with random
neighbour particle hopping is studied\cite{lubeck}. Quenched averages 
for self avoiding walks on random lattices\cite{dd1}, 
asymptotic shape of the region
visited by an Eulerian walker\cite{dd2}, linear and branched avalanches
are studied in self avoiding random walks\cite{manna}, effect of 
quenching is studied in quantum random walk\cite{ps}. The drift and
trapping in biased diffusion on disordered lattices
is also studied\cite{stauffer}.

Recently, some more interesting results on random walk are repoted.
The average number of distinct sites visited by a random walker on the 
random graph\cite{satyada1}, statistics of first passage time of the 
Browian motion conditioned by maximum value of area\cite{satyada2}
are studied recently.
It may be mentioned here that the first passage time in complex 
scale invariant media was studied very recently\cite{tejedor1}.
The thoery of mean first passage time for jump processes are developed
\cite{tejedor2} and
verified by applying in Levy flights and fractional Brownian motion.
The statistics of the gap and time interval between the highest positions
of a Markovian one dimensional random walker\cite{satyada3}, the universal
statistics of longest lasting records random walks and Levy flights
are also studied\cite{satyada4} recently.

In real life, the random walk problem has been generalised in continuum.
The exact solution of a Brownian inchworm model and self-propulsion
was also studied\cite{sriram}, theory of continuum random walks and 
application in chemotaxis was developed\cite{cont1}. 
Random walks in continuum was
also studied for diffusion and reaction in catalyst\cite{cont2}. 

The Pearson walk\cite{pearson}, is a variant of random walk which shows many interesting
results. This is defined as the walker may choose any direction
randomly, instead of taking specified direction in lattices. 
This Pearson random walk was studied\cite{redner1,redner2} with shrinking stepsize.
Very recently, the Pearson walk\cite{pearson} is studied with uniformly
distributed random size of flight\cite{aba}. The statistics of a tired Pearson
walker was also studied recently\cite{trw} to analyse the 
exit probability and first
passage time\cite{fpbook}.

In the literature of mathematics\cite{lazy1,lazy2}, the lazy 
random walk is defined as the walker having 50$\%$ chance
to move from any site and studied extensively on the lattice. 
The lazy random walk is not merely a pedagogical concept. It is already
used to study the superpixel segmentation\cite{ieee}.
{\it What will happen if a Pearson
walker becomes lazy where it's moves are probabilistic ?} In this article,
the motion of a lazy Pearson walker is studied by computer simulation and the
numerical results are reported. In the next section
(section-II), the model of lazy Pearson
walker is described and the numerical results are given. The paper ends with
a summary in section-III.
\vskip 1cm

\noindent {\bf II. Model and Results:}

The lazy random walk\cite{lazy1,lazy2} is usually described on the lattice where the walker
has 50$\%$ chance to move from any given site. In this paper, the lazy 
Pearson random walk
is described with various values of probability ($p$)
of jump from the present position, instead of $p=1/2$ as defined 
on the lattice. In two dimension, a lazy Pearson walker starts 
its journey from the
origin and jumps (unit distance)  with probability $p$ 
in any direction ($\theta$) chosen randomly (unformly distributed) between
0 and $2\pi$. In two dimensions, the rule of the jump of the lazy Pearson walker
may be expressed by following Markovian evolution:

\begin{center} 
$x(t+1) = x(t) + {\rm cos} \theta$\\
$y(t+1) = y(t) + {\rm sin} \theta$\end{center}

The exit probability ($P_e$) of a lazy walker is defined as the probability of
exit (first time) of a walker from a circular/spherical (in 2D/3D
respectively) zone 
specified by its radius $r_e$, in a given 
time of observation $N_t$. This probability
is calculated here over $N_s$ number of different random samples.

Figure-1(a) shows such a plot of exit probability ($P_e$) as a function
of radius ($r_e$) of exit zone for different values of probability ($p$)
of jump of a lazy Pearson walker in two dimensions. For a given value of
$p$, the exit probability ($P_e$) decreases as the radius ($r_e$) of exit
zone increases, in a given time of observation ($N_t = 10^4$ here). As
the probability of jump ($p$) decreases, the exit probability ($P_e$) falls
in a faster rate as $r_e$ increases. A careful inspection shows that 
for a fixed value of $p$, the 
$P_e$ is almost constant upto a certain value of $r_e$ and then decreases
monotonically. Further, it may be noted that for a given value of $r_e$, the
exit probability $P_e$ decreases as $p$ decreases. These observation promted
to assume a scaling like 
${P_e}p^{\alpha}=F({r_e}p^{\beta})$, where $\alpha$, $\beta$ are some numbers
and $F$ is a function. The curves represented by the different symbols
(different values of $p$) in Figure-1(a) falls in a single curve (shown in
Figure-1(b)) if one choose, $\alpha$=0 and $\beta=-1/2$. It may be mentioned 
here that the statistics is based on $N_s=5\times10^5$ number of different 
random samples in two dimensions. 

The time required by a lazy walker to exit first from the specified zone,
is called first passage time ($t_1$). 
The probability distribution ($P(t_1)$) of the first passage time is studied
for various values of probability ($p$) of jump of a lazy walker. Figure-2(a)
shows the probability distribution of first passage time for different values
of $p$. It is an unimodal function. 
Here, it may be noted that as $p$ increases, the mode of the distribution
shifts towards the lower values of $t_1$ and the distribution gets sharper
and sharper. Here also, one may think of a scaling behaviour of $P(t_1)$ as:
${P(t_1)}p^{\gamma} = G({t_1}p^{\delta})$. Using $\gamma=-1$ and $\delta=1$
the data for various values of $p$ collapse supporting the proposed
scaling behavior.
This is shown in Figure-2(b). It may be mentioned here that this scaling
behaviour is independent of the choice of $r_e$.
 
Lazy Pearson walk in three dimensions is also studied. Here, the
dynamical equations (or the algorithm of movement) may be expressed as:

\begin{center}

$x(t+1) = x(t) + {\rm sin} \theta {\rm cos} \phi$\\
$y(t+1) = y(t) + {\rm sin} \theta {\rm sin} \phi$\\
$z(t+1) = z(t) + {\rm cos} \theta$ \end{center}

Here $\theta$ is chosen randomly (uniformly distributed) between 0 and $\pi$.
$\phi$ is chosen randomly (uniformly distributed) between 0 and $2\pi$. In this
case, the exit probability ($P_e$) is studied as a function of the radius 
($r_e$) of the spherical zone for different values of probability ($p$) of
jump of a lazy walker. 
In three dimensions, the time of observation is $N_t=10^5$ and the statistics
is based on $N_s=10^5$ number of different random samples.
This is shown in Figure-3(a). The behavious are quite
similar to that observed in two dimensions (shown in Figure-1(a)). Here also,
one may think of a scaling behaviour like:
${P_e}p^{\alpha}=F({r_e}p^{\beta})$. By choosing $\alpha=0$ and $\beta=-1/2$
a fair data collapse is obtained which supports the assumed scaling behaviour. 
This is shown in Figure-3(b). 

The probability distribution ($P(t_1)$) of first passage time ($t_1$) 
of a lazy Pearson walker is
also studied in three dimensions for different values of 
probability ($p$) of jump and shown in Figure-4(a). The variations are
quite similar to that observed in two dimensional lazy walker. A scaling
like, 
${P(t_1)}p^{\gamma} = G({t_1}p^{\delta})$, is proposed here. Choosing
$\gamma=-1$ and $\delta=1$, this scaling behaviour of the probability
distribution of first passage time was established numerically by the
method of data collapse. This is shown in Figure-4(b). Here also, it
is observed that this scaling behaviour is independent of the choice of $r_e$.

\vskip 1cm
 
\noindent {\bf III. Summary:}

In this paper, 
the motion of a lazy Pearson walker is studied
with different probability ($p$) of jump in two and three dimensions,
by computer simulation. The exit probability and the 
probability distribution of first passage time are studied. The probability of exit 
($P_e$) from a zone of
radius $r_e$, is studied as a function of $r_e$ with different values of
jump probability $p$. 
Here, $p$ can take any value between 0 and 1, unlike the case of conventional
lazy walker. For a given value of $p$, the exit probability was found to fall
as $r_e$ grows.
The exit probability $P_e$ is found to scale as
${P_e}p^{\alpha}=F({r_e}p^{\beta})$, 
which is obtained by 
method of data collapse. 

The first passage time ($t_1$) i.e., the time 
required for first exit from a zone is studied. The probability distribution
($P(t_1)$) of first passage time was studied for different values of jump
probability ($p$). 
The probability distribution of first passage time, is a nonmonotonic unimodal
function. The mode serves the role of the scale of time of exit from the zone of radius
$r_e$. This time scale decreases as the probability $p$ (of jump) increase, which is
quite natural. 
The probability distribution of first passage time 
was found to scale as 
${P(t_1)}p^{\gamma} = G({t_1}p^{\delta})$. 
Where,
$F$ and $G$ are two scaling functions and $\alpha$, $\beta$, $\gamma$ and 
$\delta$ are some exponents. In both the dimensions, it is found that,
$\alpha = 0$, $\beta=-1/2$, $\gamma=-1$ and $\delta=1$. Interestingly, it was observed
that this scaling behaviour (and the exponents also) is independent of the choice
of $r_e$.

\vskip 1cm

\noindent {\bf Acknowledgements:} The library facility provided by Calcutta
University is gratefully acknowledged.

\newpage
% GNUPLOT: LaTeX picture P(exit) vs r_exit & scaling in 2D
\setlength{\unitlength}{0.240900pt}
\ifx\plotpoint\undefined\newsavebox{\plotpoint}\fi
\sbox{\plotpoint}{\rule[-0.200pt]{0.400pt}{0.400pt}}%
\begin{picture}(1125,900)(0,0)
\sbox{\plotpoint}{\rule[-0.200pt]{0.400pt}{0.400pt}}%
\put(171.0,131.0){\rule[-0.200pt]{4.818pt}{0.400pt}}
\put(151,131){\makebox(0,0)[r]{ 0}}
\put(1044.0,131.0){\rule[-0.200pt]{4.818pt}{0.400pt}}
\put(171.0,204.0){\rule[-0.200pt]{4.818pt}{0.400pt}}
\put(151,204){\makebox(0,0)[r]{ 0.1}}
\put(1044.0,204.0){\rule[-0.200pt]{4.818pt}{0.400pt}}
\put(171.0,277.0){\rule[-0.200pt]{4.818pt}{0.400pt}}
\put(151,277){\makebox(0,0)[r]{ 0.2}}
\put(1044.0,277.0){\rule[-0.200pt]{4.818pt}{0.400pt}}
\put(171.0,349.0){\rule[-0.200pt]{4.818pt}{0.400pt}}
\put(151,349){\makebox(0,0)[r]{ 0.3}}
\put(1044.0,349.0){\rule[-0.200pt]{4.818pt}{0.400pt}}
\put(171.0,422.0){\rule[-0.200pt]{4.818pt}{0.400pt}}
\put(151,422){\makebox(0,0)[r]{ 0.4}}
\put(1044.0,422.0){\rule[-0.200pt]{4.818pt}{0.400pt}}
\put(171.0,495.0){\rule[-0.200pt]{4.818pt}{0.400pt}}
\put(151,495){\makebox(0,0)[r]{ 0.5}}
\put(1044.0,495.0){\rule[-0.200pt]{4.818pt}{0.400pt}}
\put(171.0,568.0){\rule[-0.200pt]{4.818pt}{0.400pt}}
\put(151,568){\makebox(0,0)[r]{ 0.6}}
\put(1044.0,568.0){\rule[-0.200pt]{4.818pt}{0.400pt}}
\put(171.0,641.0){\rule[-0.200pt]{4.818pt}{0.400pt}}
\put(151,641){\makebox(0,0)[r]{ 0.7}}
\put(1044.0,641.0){\rule[-0.200pt]{4.818pt}{0.400pt}}
\put(171.0,713.0){\rule[-0.200pt]{4.818pt}{0.400pt}}
\put(151,713){\makebox(0,0)[r]{ 0.8}}
\put(1044.0,713.0){\rule[-0.200pt]{4.818pt}{0.400pt}}
\put(171.0,786.0){\rule[-0.200pt]{4.818pt}{0.400pt}}
\put(151,786){\makebox(0,0)[r]{ 0.9}}
\put(1044.0,786.0){\rule[-0.200pt]{4.818pt}{0.400pt}}
\put(171.0,859.0){\rule[-0.200pt]{4.818pt}{0.400pt}}
\put(151,859){\makebox(0,0)[r]{ 1}}
\put(1044.0,859.0){\rule[-0.200pt]{4.818pt}{0.400pt}}
\put(171.0,131.0){\rule[-0.200pt]{0.400pt}{4.818pt}}
\put(171,90){\makebox(0,0){ 0}}
\put(171.0,839.0){\rule[-0.200pt]{0.400pt}{4.818pt}}
\put(350.0,131.0){\rule[-0.200pt]{0.400pt}{4.818pt}}
\put(350,90){\makebox(0,0){ 20}}
\put(350.0,839.0){\rule[-0.200pt]{0.400pt}{4.818pt}}
\put(528.0,131.0){\rule[-0.200pt]{0.400pt}{4.818pt}}
\put(528,90){\makebox(0,0){ 40}}
\put(528.0,839.0){\rule[-0.200pt]{0.400pt}{4.818pt}}
\put(707.0,131.0){\rule[-0.200pt]{0.400pt}{4.818pt}}
\put(707,90){\makebox(0,0){ 60}}
\put(707.0,839.0){\rule[-0.200pt]{0.400pt}{4.818pt}}
\put(885.0,131.0){\rule[-0.200pt]{0.400pt}{4.818pt}}
\put(885,90){\makebox(0,0){ 80}}
\put(885.0,839.0){\rule[-0.200pt]{0.400pt}{4.818pt}}
\put(1064.0,131.0){\rule[-0.200pt]{0.400pt}{4.818pt}}
\put(1064,90){\makebox(0,0){ 100}}
\put(1064.0,839.0){\rule[-0.200pt]{0.400pt}{4.818pt}}
\put(171.0,131.0){\rule[-0.200pt]{0.400pt}{175.375pt}}
\put(171.0,131.0){\rule[-0.200pt]{215.124pt}{0.400pt}}
\put(1064.0,131.0){\rule[-0.200pt]{0.400pt}{175.375pt}}
\put(171.0,859.0){\rule[-0.200pt]{215.124pt}{0.400pt}}
\put(30,495){\makebox(0,0){$P_e$}}
\put(617,29){\makebox(0,0){$r_e$}}
\put(975,786){\makebox(0,0)[l]{(a)}}
\put(216,859){\makebox(0,0){$\blacksquare$}}
\put(260,859){\makebox(0,0){$\blacksquare$}}
\put(305,859){\makebox(0,0){$\blacksquare$}}
\put(350,858){\makebox(0,0){$\blacksquare$}}
\put(394,846){\makebox(0,0){$\blacksquare$}}
\put(439,809){\makebox(0,0){$\blacksquare$}}
\put(484,743){\makebox(0,0){$\blacksquare$}}
\put(528,662){\makebox(0,0){$\blacksquare$}}
\put(573,573){\makebox(0,0){$\blacksquare$}}
\put(618,488){\makebox(0,0){$\blacksquare$}}
\put(662,410){\makebox(0,0){$\blacksquare$}}
\put(707,343){\makebox(0,0){$\blacksquare$}}
\put(751,288){\makebox(0,0){$\blacksquare$}}
\put(796,244){\makebox(0,0){$\blacksquare$}}
\put(841,210){\makebox(0,0){$\blacksquare$}}
\put(885,185){\makebox(0,0){$\blacksquare$}}
\put(930,167){\makebox(0,0){$\blacksquare$}}
\put(975,154){\makebox(0,0){$\blacksquare$}}
\put(1019,146){\makebox(0,0){$\blacksquare$}}
\put(1064,140){\makebox(0,0){$\blacksquare$}}
\put(216,859){\makebox(0,0){$\bullet$}}
\put(260,859){\makebox(0,0){$\bullet$}}
\put(305,859){\makebox(0,0){$\bullet$}}
\put(350,859){\makebox(0,0){$\bullet$}}
\put(394,859){\makebox(0,0){$\bullet$}}
\put(439,857){\makebox(0,0){$\bullet$}}
\put(484,848){\makebox(0,0){$\bullet$}}
\put(528,826){\makebox(0,0){$\bullet$}}
\put(573,789){\makebox(0,0){$\bullet$}}
\put(618,740){\makebox(0,0){$\bullet$}}
\put(662,681){\makebox(0,0){$\bullet$}}
\put(707,620){\makebox(0,0){$\bullet$}}
\put(751,557){\makebox(0,0){$\bullet$}}
\put(796,497){\makebox(0,0){$\bullet$}}
\put(841,440){\makebox(0,0){$\bullet$}}
\put(885,389){\makebox(0,0){$\bullet$}}
\put(930,343){\makebox(0,0){$\bullet$}}
\put(975,304){\makebox(0,0){$\bullet$}}
\put(1019,269){\makebox(0,0){$\bullet$}}
\put(1064,240){\makebox(0,0){$\bullet$}}
\sbox{\plotpoint}{\rule[-0.400pt]{0.800pt}{0.800pt}}%
\put(216,859){\makebox(0,0){$\blacktriangle$}}
\put(260,859){\makebox(0,0){$\blacktriangle$}}
\put(305,859){\makebox(0,0){$\blacktriangle$}}
\put(350,859){\makebox(0,0){$\blacktriangle$}}
\put(394,859){\makebox(0,0){$\blacktriangle$}}
\put(439,859){\makebox(0,0){$\blacktriangle$}}
\put(484,858){\makebox(0,0){$\blacktriangle$}}
\put(528,853){\makebox(0,0){$\blacktriangle$}}
\put(573,842){\makebox(0,0){$\blacktriangle$}}
\put(618,821){\makebox(0,0){$\blacktriangle$}}
\put(662,790){\makebox(0,0){$\blacktriangle$}}
\put(707,751){\makebox(0,0){$\blacktriangle$}}
\put(751,706){\makebox(0,0){$\blacktriangle$}}
\put(796,658){\makebox(0,0){$\blacktriangle$}}
\put(841,606){\makebox(0,0){$\blacktriangle$}}
\put(885,556){\makebox(0,0){$\blacktriangle$}}
\put(930,506){\makebox(0,0){$\blacktriangle$}}
\put(975,460){\makebox(0,0){$\blacktriangle$}}
\put(1019,416){\makebox(0,0){$\blacktriangle$}}
\put(1064,375){\makebox(0,0){$\blacktriangle$}}
\sbox{\plotpoint}{\rule[-0.500pt]{1.000pt}{1.000pt}}%
\put(216,859){\makebox(0,0){$\blacktriangledown$}}
\put(260,859){\makebox(0,0){$\blacktriangledown$}}
\put(305,859){\makebox(0,0){$\blacktriangledown$}}
\put(350,859){\makebox(0,0){$\blacktriangledown$}}
\put(394,859){\makebox(0,0){$\blacktriangledown$}}
\put(439,859){\makebox(0,0){$\blacktriangledown$}}
\put(484,859){\makebox(0,0){$\blacktriangledown$}}
\put(528,858){\makebox(0,0){$\blacktriangledown$}}
\put(573,855){\makebox(0,0){$\blacktriangledown$}}
\put(618,847){\makebox(0,0){$\blacktriangledown$}}
\put(662,832){\makebox(0,0){$\blacktriangledown$}}
\put(707,810){\makebox(0,0){$\blacktriangledown$}}
\put(751,782){\makebox(0,0){$\blacktriangledown$}}
\put(796,747){\makebox(0,0){$\blacktriangledown$}}
\put(841,708){\makebox(0,0){$\blacktriangledown$}}
\put(885,666){\makebox(0,0){$\blacktriangledown$}}
\put(930,621){\makebox(0,0){$\blacktriangledown$}}
\put(975,578){\makebox(0,0){$\blacktriangledown$}}
\put(1019,534){\makebox(0,0){$\blacktriangledown$}}
\put(1064,492){\makebox(0,0){$\blacktriangledown$}}
\sbox{\plotpoint}{\rule[-0.200pt]{0.400pt}{0.400pt}}%
\put(171.0,131.0){\rule[-0.200pt]{0.400pt}{175.375pt}}
\put(171.0,131.0){\rule[-0.200pt]{215.124pt}{0.400pt}}
\put(1064.0,131.0){\rule[-0.200pt]{0.400pt}{175.375pt}}
\put(171.0,859.0){\rule[-0.200pt]{215.124pt}{0.400pt}}
\end{picture}

\begin{picture}(1125,900)(0,0)
\put(171.0,131.0){\rule[-0.200pt]{4.818pt}{0.400pt}}
\put(151,131){\makebox(0,0)[r]{ 0}}
\put(1044.0,131.0){\rule[-0.200pt]{4.818pt}{0.400pt}}
\put(171.0,204.0){\rule[-0.200pt]{4.818pt}{0.400pt}}
\put(151,204){\makebox(0,0)[r]{ 0.1}}
\put(1044.0,204.0){\rule[-0.200pt]{4.818pt}{0.400pt}}
\put(171.0,277.0){\rule[-0.200pt]{4.818pt}{0.400pt}}
\put(151,277){\makebox(0,0)[r]{ 0.2}}
\put(1044.0,277.0){\rule[-0.200pt]{4.818pt}{0.400pt}}
\put(171.0,349.0){\rule[-0.200pt]{4.818pt}{0.400pt}}
\put(151,349){\makebox(0,0)[r]{ 0.3}}
\put(1044.0,349.0){\rule[-0.200pt]{4.818pt}{0.400pt}}
\put(171.0,422.0){\rule[-0.200pt]{4.818pt}{0.400pt}}
\put(151,422){\makebox(0,0)[r]{ 0.4}}
\put(1044.0,422.0){\rule[-0.200pt]{4.818pt}{0.400pt}}
\put(171.0,495.0){\rule[-0.200pt]{4.818pt}{0.400pt}}
\put(151,495){\makebox(0,0)[r]{ 0.5}}
\put(1044.0,495.0){\rule[-0.200pt]{4.818pt}{0.400pt}}
\put(171.0,568.0){\rule[-0.200pt]{4.818pt}{0.400pt}}
\put(151,568){\makebox(0,0)[r]{ 0.6}}
\put(1044.0,568.0){\rule[-0.200pt]{4.818pt}{0.400pt}}
\put(171.0,641.0){\rule[-0.200pt]{4.818pt}{0.400pt}}
\put(151,641){\makebox(0,0)[r]{ 0.7}}
\put(1044.0,641.0){\rule[-0.200pt]{4.818pt}{0.400pt}}
\put(171.0,713.0){\rule[-0.200pt]{4.818pt}{0.400pt}}
\put(151,713){\makebox(0,0)[r]{ 0.8}}
\put(1044.0,713.0){\rule[-0.200pt]{4.818pt}{0.400pt}}
\put(171.0,786.0){\rule[-0.200pt]{4.818pt}{0.400pt}}
\put(151,786){\makebox(0,0)[r]{ 0.9}}
\put(1044.0,786.0){\rule[-0.200pt]{4.818pt}{0.400pt}}
\put(171.0,859.0){\rule[-0.200pt]{4.818pt}{0.400pt}}
\put(151,859){\makebox(0,0)[r]{ 1}}
\put(1044.0,859.0){\rule[-0.200pt]{4.818pt}{0.400pt}}
\put(171.0,131.0){\rule[-0.200pt]{0.400pt}{4.818pt}}
\put(171,90){\makebox(0,0){ 0}}
\put(171.0,839.0){\rule[-0.200pt]{0.400pt}{4.818pt}}
\put(350.0,131.0){\rule[-0.200pt]{0.400pt}{4.818pt}}
\put(350,90){\makebox(0,0){ 50}}
\put(350.0,839.0){\rule[-0.200pt]{0.400pt}{4.818pt}}
\put(528.0,131.0){\rule[-0.200pt]{0.400pt}{4.818pt}}
\put(528,90){\makebox(0,0){ 100}}
\put(528.0,839.0){\rule[-0.200pt]{0.400pt}{4.818pt}}
\put(707.0,131.0){\rule[-0.200pt]{0.400pt}{4.818pt}}
\put(707,90){\makebox(0,0){ 150}}
\put(707.0,839.0){\rule[-0.200pt]{0.400pt}{4.818pt}}
\put(885.0,131.0){\rule[-0.200pt]{0.400pt}{4.818pt}}
\put(885,90){\makebox(0,0){ 200}}
\put(885.0,839.0){\rule[-0.200pt]{0.400pt}{4.818pt}}
\put(1064.0,131.0){\rule[-0.200pt]{0.400pt}{4.818pt}}
\put(1064,90){\makebox(0,0){ 250}}
\put(1064.0,839.0){\rule[-0.200pt]{0.400pt}{4.818pt}}
\put(171.0,131.0){\rule[-0.200pt]{0.400pt}{175.375pt}}
\put(171.0,131.0){\rule[-0.200pt]{215.124pt}{0.400pt}}
\put(1064.0,131.0){\rule[-0.200pt]{0.400pt}{175.375pt}}
\put(171.0,859.0){\rule[-0.200pt]{215.124pt}{0.400pt}}
\put(30,495){\makebox(0,0){${P_e}p^{\alpha}$}}
\put(617,29){\makebox(0,0){${r_e}p^{\beta}$}}
\put(778,713){\makebox(0,0)[l]{$\beta=-0.5$}}
\put(778,788){\makebox(0,0)[l]{$\alpha=0.0$}}
\put(993,786){\makebox(0,0)[l]{(b)}}
\put(211,859){\makebox(0,0){$\blacksquare$}}
\put(251,859){\makebox(0,0){$\blacksquare$}}
\put(291,859){\makebox(0,0){$\blacksquare$}}
\put(331,858){\makebox(0,0){$\blacksquare$}}
\put(371,846){\makebox(0,0){$\blacksquare$}}
\put(411,809){\makebox(0,0){$\blacksquare$}}
\put(451,743){\makebox(0,0){$\blacksquare$}}
\put(490,662){\makebox(0,0){$\blacksquare$}}
\put(530,573){\makebox(0,0){$\blacksquare$}}
\put(570,488){\makebox(0,0){$\blacksquare$}}
\put(610,410){\makebox(0,0){$\blacksquare$}}
\put(650,343){\makebox(0,0){$\blacksquare$}}
\put(690,288){\makebox(0,0){$\blacksquare$}}
\put(730,244){\makebox(0,0){$\blacksquare$}}
\put(770,210){\makebox(0,0){$\blacksquare$}}
\put(810,185){\makebox(0,0){$\blacksquare$}}
\put(850,167){\makebox(0,0){$\blacksquare$}}
\put(890,154){\makebox(0,0){$\blacksquare$}}
\put(930,146){\makebox(0,0){$\blacksquare$}}
\put(970,140){\makebox(0,0){$\blacksquare$}}
\put(199,859){\makebox(0,0){$\bullet$}}
\put(227,859){\makebox(0,0){$\bullet$}}
\put(256,859){\makebox(0,0){$\bullet$}}
\put(284,859){\makebox(0,0){$\bullet$}}
\put(312,859){\makebox(0,0){$\bullet$}}
\put(340,857){\makebox(0,0){$\bullet$}}
\put(369,848){\makebox(0,0){$\bullet$}}
\put(397,826){\makebox(0,0){$\bullet$}}
\put(425,789){\makebox(0,0){$\bullet$}}
\put(453,740){\makebox(0,0){$\bullet$}}
\put(482,681){\makebox(0,0){$\bullet$}}
\put(510,620){\makebox(0,0){$\bullet$}}
\put(538,557){\makebox(0,0){$\bullet$}}
\put(566,497){\makebox(0,0){$\bullet$}}
\put(595,440){\makebox(0,0){$\bullet$}}
\put(623,389){\makebox(0,0){$\bullet$}}
\put(651,343){\makebox(0,0){$\bullet$}}
\put(679,304){\makebox(0,0){$\bullet$}}
\put(708,269){\makebox(0,0){$\bullet$}}
\put(736,240){\makebox(0,0){$\bullet$}}
\sbox{\plotpoint}{\rule[-0.400pt]{0.800pt}{0.800pt}}%
\put(194,859){\makebox(0,0){$\blacktriangle$}}
\put(217,859){\makebox(0,0){$\blacktriangle$}}
\put(240,859){\makebox(0,0){$\blacktriangle$}}
\put(263,859){\makebox(0,0){$\blacktriangle$}}
\put(286,859){\makebox(0,0){$\blacktriangle$}}
\put(309,859){\makebox(0,0){$\blacktriangle$}}
\put(332,858){\makebox(0,0){$\blacktriangle$}}
\put(355,853){\makebox(0,0){$\blacktriangle$}}
\put(379,842){\makebox(0,0){$\blacktriangle$}}
\put(402,821){\makebox(0,0){$\blacktriangle$}}
\put(425,790){\makebox(0,0){$\blacktriangle$}}
\put(448,751){\makebox(0,0){$\blacktriangle$}}
\put(471,706){\makebox(0,0){$\blacktriangle$}}
\put(494,658){\makebox(0,0){$\blacktriangle$}}
\put(517,606){\makebox(0,0){$\blacktriangle$}}
\put(540,556){\makebox(0,0){$\blacktriangle$}}
\put(563,506){\makebox(0,0){$\blacktriangle$}}
\put(586,460){\makebox(0,0){$\blacktriangle$}}
\put(609,416){\makebox(0,0){$\blacktriangle$}}
\put(632,375){\makebox(0,0){$\blacktriangle$}}
\sbox{\plotpoint}{\rule[-0.500pt]{1.000pt}{1.000pt}}%
\put(191,859){\makebox(0,0){$\blacktriangledown$}}
\put(211,859){\makebox(0,0){$\blacktriangledown$}}
\put(231,859){\makebox(0,0){$\blacktriangledown$}}
\put(251,859){\makebox(0,0){$\blacktriangledown$}}
\put(271,859){\makebox(0,0){$\blacktriangledown$}}
\put(291,859){\makebox(0,0){$\blacktriangledown$}}
\put(311,859){\makebox(0,0){$\blacktriangledown$}}
\put(331,858){\makebox(0,0){$\blacktriangledown$}}
\put(351,855){\makebox(0,0){$\blacktriangledown$}}
\put(371,847){\makebox(0,0){$\blacktriangledown$}}
\put(391,832){\makebox(0,0){$\blacktriangledown$}}
\put(411,810){\makebox(0,0){$\blacktriangledown$}}
\put(431,782){\makebox(0,0){$\blacktriangledown$}}
\put(451,747){\makebox(0,0){$\blacktriangledown$}}
\put(471,708){\makebox(0,0){$\blacktriangledown$}}
\put(490,666){\makebox(0,0){$\blacktriangledown$}}
\put(510,621){\makebox(0,0){$\blacktriangledown$}}
\put(530,578){\makebox(0,0){$\blacktriangledown$}}
\put(550,534){\makebox(0,0){$\blacktriangledown$}}
\put(570,492){\makebox(0,0){$\blacktriangledown$}}
\sbox{\plotpoint}{\rule[-0.200pt]{0.400pt}{0.400pt}}%
\put(171.0,131.0){\rule[-0.200pt]{0.400pt}{175.375pt}}
\put(171.0,131.0){\rule[-0.200pt]{215.124pt}{0.400pt}}
\put(1064.0,131.0){\rule[-0.200pt]{0.400pt}{175.375pt}}
\put(171.0,859.0){\rule[-0.200pt]{215.124pt}{0.400pt}}
\end{picture}

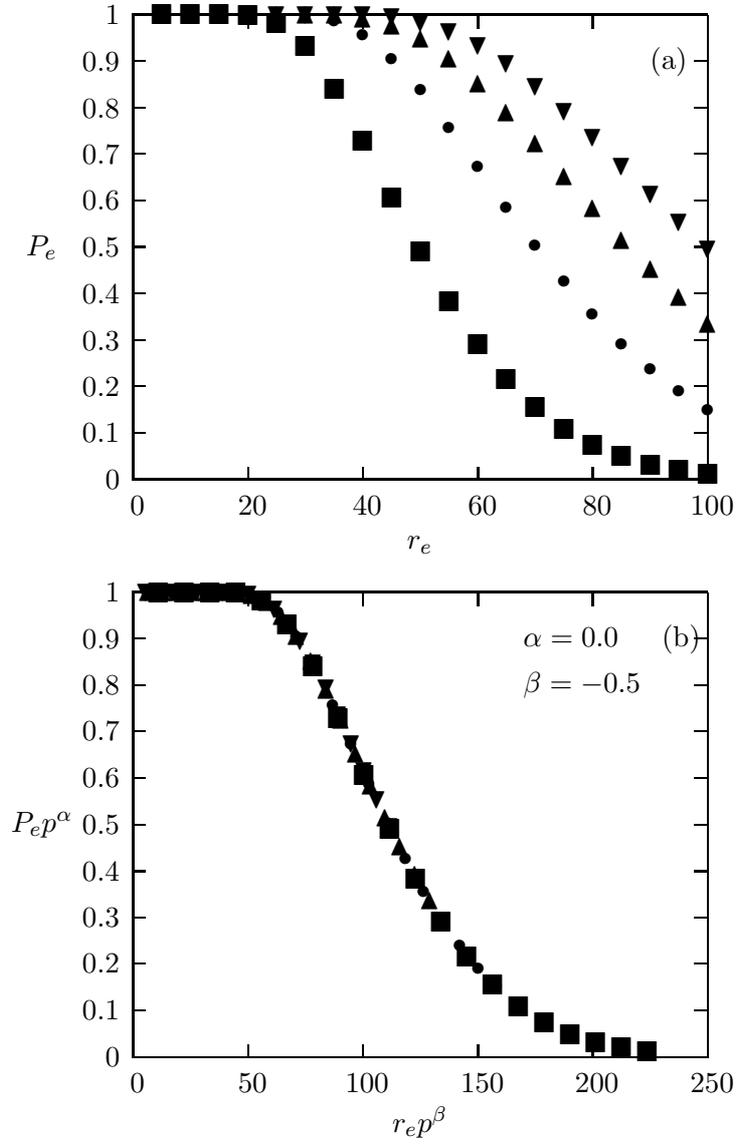
\noindent {\bf Fig-1.} The exit probability ($P_e$) versus exit
radius ($r_e$) (a) and
its scaling (b). Different symbol correspond to different values of jump
probability ($p$) in two dimensions. $p=0.2$($\blacksquare$), $p=0.4(\bullet)$, 
$p=0.6(\blacktriangle$) and $p=0.8(\blacktriangledown)$. 
Here, $N_s = 5\times 10^5$ and $N_t=10^4$.

\newpage

%%%%FIGURE Distribution of First passage time/scaling in 2D
% GNUPLOT: LaTeX picture
\setlength{\unitlength}{0.240900pt}
\ifx\plotpoint\undefined\newsavebox{\plotpoint}\fi
\sbox{\plotpoint}{\rule[-0.200pt]{0.400pt}{0.400pt}}%


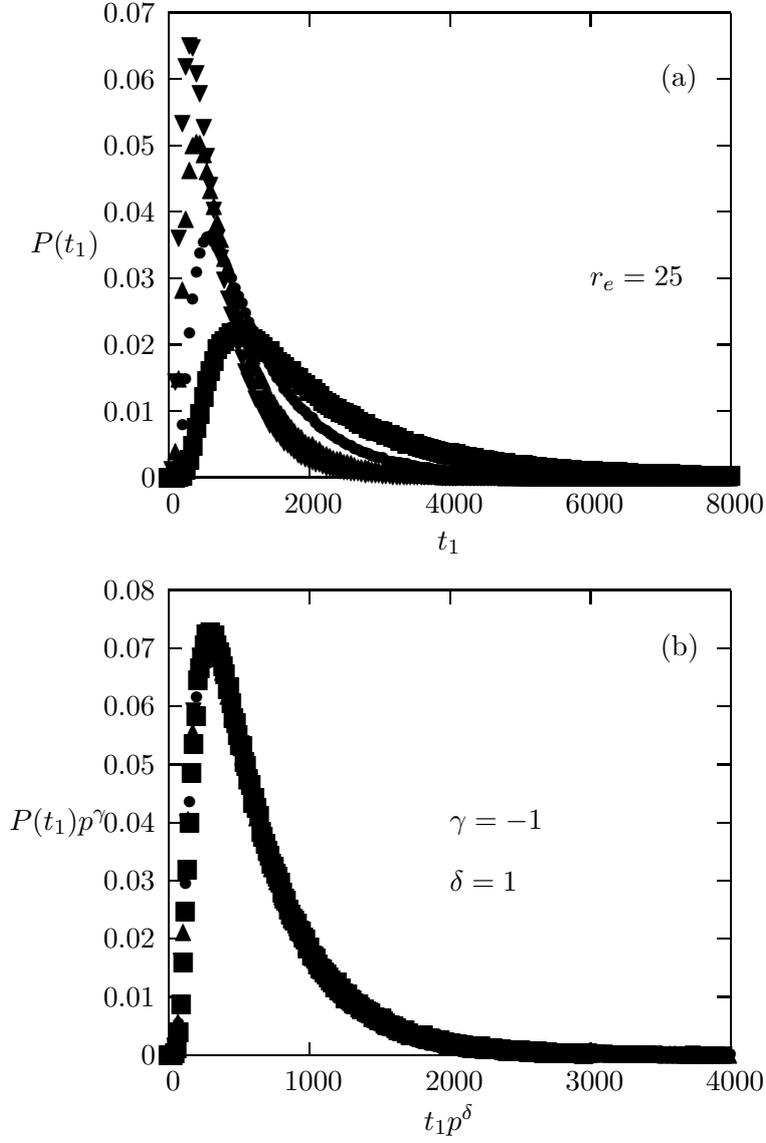
\noindent {\bf Fig-2.} The distribution (a) of first passage time ($t_1$) and
its scaling (b). Different symbol correspond to different values of jump
probability ($p$) in two dimensions. $p=0.3$($\blacksquare$), $p=0.5(\bullet)$, $p=0.7(\blacktriangle$) and $p=0.9(\blacktriangledown)$.
Here, $N_s = 5\times10^5$, $N_t=10^4$ and $r_e=25$.
 
\newpage
% GNUPLOT: LaTeX picture P(exit) versus r_exit in 3D
\setlength{\unitlength}{0.240900pt}
\ifx\plotpoint\undefined\newsavebox{\plotpoint}\fi
\sbox{\plotpoint}{\rule[-0.200pt]{0.400pt}{0.400pt}}%
%

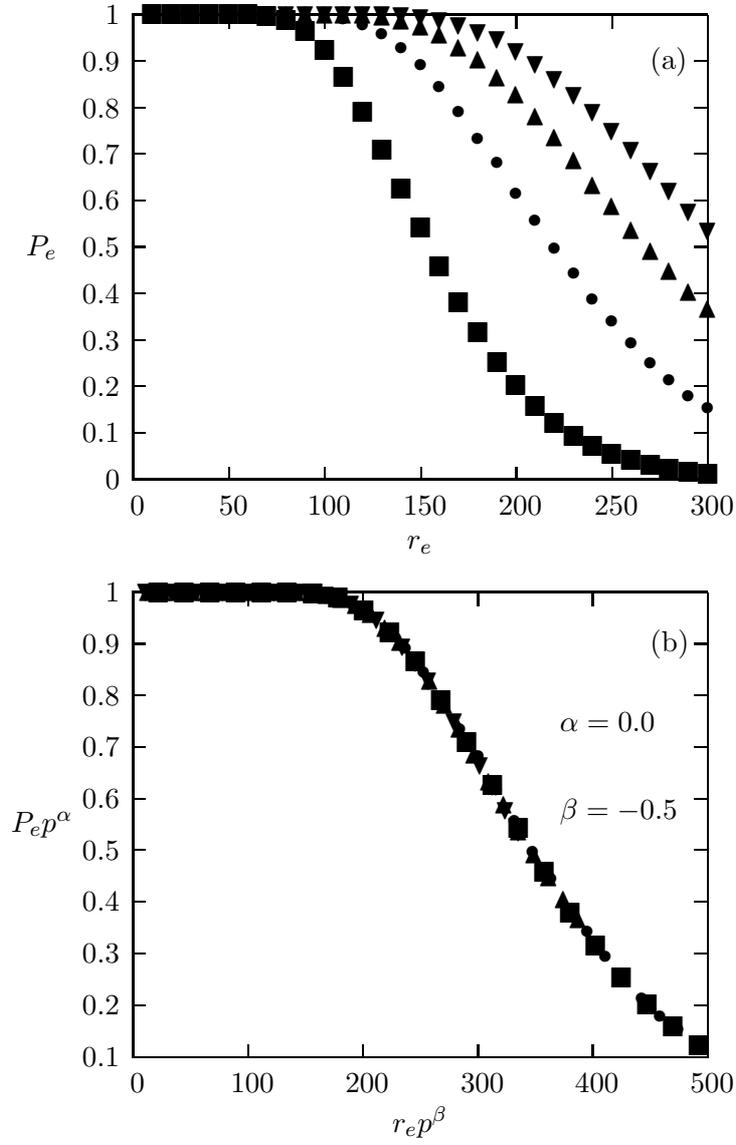
\noindent {\bf Fig-3.} The exit probability ($P_e$) versus exit
radius ($r_e$) (a) and
its scaling (b). Different symbol correspond to different values of jump
probability ($p$) in three dimensions. $p=0.2$($\blacksquare$), $p=0.4(\bullet)$, 
$p=0.6(\blacktriangle$) and $p=0.8(\blacktriangledown)$. 
Here, $N_s = 10^5$ and $N_t=10^5$.

\newpage
% Distribution of first passage time and its scaling in 3D
% GNUPLOT: LaTeX picture
\setlength{\unitlength}{0.240900pt}
\ifx\plotpoint\undefined\newsavebox{\plotpoint}\fi
\sbox{\plotpoint}{\rule[-0.200pt]{0.400pt}{0.400pt}}%
\begin{picture}(1125,900)(0,0)
\sbox{\plotpoint}{\rule[-0.200pt]{0.400pt}{0.400pt}}%
\put(191.0,131.0){\rule[-0.200pt]{4.818pt}{0.400pt}}
\put(171,131){\makebox(0,0)[r]{ 0}}
\put(1044.0,131.0){\rule[-0.200pt]{4.818pt}{0.400pt}}
\put(191.0,235.0){\rule[-0.200pt]{4.818pt}{0.400pt}}
\put(171,235){\makebox(0,0)[r]{ 0.02}}
\put(1044.0,235.0){\rule[-0.200pt]{4.818pt}{0.400pt}}
\put(191.0,339.0){\rule[-0.200pt]{4.818pt}{0.400pt}}
\put(171,339){\makebox(0,0)[r]{ 0.04}}
\put(1044.0,339.0){\rule[-0.200pt]{4.818pt}{0.400pt}}
\put(191.0,443.0){\rule[-0.200pt]{4.818pt}{0.400pt}}
\put(171,443){\makebox(0,0)[r]{ 0.06}}
\put(1044.0,443.0){\rule[-0.200pt]{4.818pt}{0.400pt}}
\put(191.0,547.0){\rule[-0.200pt]{4.818pt}{0.400pt}}
\put(171,547){\makebox(0,0)[r]{ 0.08}}
\put(1044.0,547.0){\rule[-0.200pt]{4.818pt}{0.400pt}}
\put(191.0,651.0){\rule[-0.200pt]{4.818pt}{0.400pt}}
\put(171,651){\makebox(0,0)[r]{ 0.1}}
\put(1044.0,651.0){\rule[-0.200pt]{4.818pt}{0.400pt}}
\put(191.0,755.0){\rule[-0.200pt]{4.818pt}{0.400pt}}
\put(171,755){\makebox(0,0)[r]{ 0.12}}
\put(1044.0,755.0){\rule[-0.200pt]{4.818pt}{0.400pt}}
\put(191.0,859.0){\rule[-0.200pt]{4.818pt}{0.400pt}}
\put(171,859){\makebox(0,0)[r]{ 0.14}}
\put(1044.0,859.0){\rule[-0.200pt]{4.818pt}{0.400pt}}
\put(191.0,131.0){\rule[-0.200pt]{0.400pt}{4.818pt}}
\put(191,90){\makebox(0,0){ 0}}
\put(191.0,839.0){\rule[-0.200pt]{0.400pt}{4.818pt}}
\put(409.0,131.0){\rule[-0.200pt]{0.400pt}{4.818pt}}
\put(409,90){\makebox(0,0){ 3000}}
\put(409.0,839.0){\rule[-0.200pt]{0.400pt}{4.818pt}}
\put(628.0,131.0){\rule[-0.200pt]{0.400pt}{4.818pt}}
\put(628,90){\makebox(0,0){ 6000}}
\put(628.0,839.0){\rule[-0.200pt]{0.400pt}{4.818pt}}
\put(846.0,131.0){\rule[-0.200pt]{0.400pt}{4.818pt}}
\put(846,90){\makebox(0,0){ 9000}}
\put(846.0,839.0){\rule[-0.200pt]{0.400pt}{4.818pt}}
\put(1064.0,131.0){\rule[-0.200pt]{0.400pt}{4.818pt}}
\put(1064,90){\makebox(0,0){ 12000}}
\put(1064.0,839.0){\rule[-0.200pt]{0.400pt}{4.818pt}}
\put(191.0,131.0){\rule[-0.200pt]{0.400pt}{175.375pt}}
\put(191.0,131.0){\rule[-0.200pt]{210.306pt}{0.400pt}}
\put(1064.0,131.0){\rule[-0.200pt]{0.400pt}{175.375pt}}
\put(191.0,859.0){\rule[-0.200pt]{210.306pt}{0.400pt}}
\put(30,495){\makebox(0,0){$P(t_1)$}}
\put(627,29){\makebox(0,0){$t_1$}}
\put(919,755){\makebox(0,0)[l]{(a)}}
\put(700,547){\makebox(0,0)[l]{$r_e=30$}}
\put(191,131){\makebox(0,0){$\blacksquare$}}
\put(202,131){\makebox(0,0){$\blacksquare$}}
\put(213,131){\makebox(0,0){$\blacksquare$}}
\put(224,134){\makebox(0,0){$\blacksquare$}}
\put(235,139){\makebox(0,0){$\blacksquare$}}
\put(246,152){\makebox(0,0){$\blacksquare$}}
\put(256,170){\makebox(0,0){$\blacksquare$}}
\put(267,192){\makebox(0,0){$\blacksquare$}}
\put(278,211){\makebox(0,0){$\blacksquare$}}
\put(289,232){\makebox(0,0){$\blacksquare$}}
\put(300,248){\makebox(0,0){$\blacksquare$}}
\put(311,256){\makebox(0,0){$\blacksquare$}}
\put(322,277){\makebox(0,0){$\blacksquare$}}
\put(333,282){\makebox(0,0){$\blacksquare$}}
\put(344,285){\makebox(0,0){$\blacksquare$}}
\put(355,291){\makebox(0,0){$\blacksquare$}}
\put(366,292){\makebox(0,0){$\blacksquare$}}
\put(377,293){\makebox(0,0){$\blacksquare$}}
\put(387,288){\makebox(0,0){$\blacksquare$}}
\put(398,288){\makebox(0,0){$\blacksquare$}}
\put(409,283){\makebox(0,0){$\blacksquare$}}
\put(420,280){\makebox(0,0){$\blacksquare$}}
\put(431,273){\makebox(0,0){$\blacksquare$}}
\put(442,276){\makebox(0,0){$\blacksquare$}}
\put(453,261){\makebox(0,0){$\blacksquare$}}
\put(464,259){\makebox(0,0){$\blacksquare$}}
\put(475,255){\makebox(0,0){$\blacksquare$}}
\put(486,252){\makebox(0,0){$\blacksquare$}}
\put(497,245){\makebox(0,0){$\blacksquare$}}
\put(507,236){\makebox(0,0){$\blacksquare$}}
\put(518,234){\makebox(0,0){$\blacksquare$}}
\put(529,230){\makebox(0,0){$\blacksquare$}}
\put(540,226){\makebox(0,0){$\blacksquare$}}
\put(551,223){\makebox(0,0){$\blacksquare$}}
\put(562,215){\makebox(0,0){$\blacksquare$}}
\put(573,210){\makebox(0,0){$\blacksquare$}}
\put(584,210){\makebox(0,0){$\blacksquare$}}
\put(595,202){\makebox(0,0){$\blacksquare$}}
\put(606,203){\makebox(0,0){$\blacksquare$}}
\put(617,195){\makebox(0,0){$\blacksquare$}}
\put(628,192){\makebox(0,0){$\blacksquare$}}
\put(638,190){\makebox(0,0){$\blacksquare$}}
\put(649,186){\makebox(0,0){$\blacksquare$}}
\put(660,184){\makebox(0,0){$\blacksquare$}}
\put(671,181){\makebox(0,0){$\blacksquare$}}
\put(682,182){\makebox(0,0){$\blacksquare$}}
\put(693,177){\makebox(0,0){$\blacksquare$}}
\put(704,175){\makebox(0,0){$\blacksquare$}}
\put(715,173){\makebox(0,0){$\blacksquare$}}
\put(726,172){\makebox(0,0){$\blacksquare$}}
\put(737,167){\makebox(0,0){$\blacksquare$}}
\put(748,167){\makebox(0,0){$\blacksquare$}}
\put(758,164){\makebox(0,0){$\blacksquare$}}
\put(769,161){\makebox(0,0){$\blacksquare$}}
\put(780,160){\makebox(0,0){$\blacksquare$}}
\put(791,159){\makebox(0,0){$\blacksquare$}}
\put(802,157){\makebox(0,0){$\blacksquare$}}
\put(813,155){\makebox(0,0){$\blacksquare$}}
\put(824,155){\makebox(0,0){$\blacksquare$}}
\put(835,155){\makebox(0,0){$\blacksquare$}}
\put(846,153){\makebox(0,0){$\blacksquare$}}
\put(857,152){\makebox(0,0){$\blacksquare$}}
\put(868,149){\makebox(0,0){$\blacksquare$}}
\put(878,149){\makebox(0,0){$\blacksquare$}}
\put(889,150){\makebox(0,0){$\blacksquare$}}
\put(900,148){\makebox(0,0){$\blacksquare$}}
\put(911,146){\makebox(0,0){$\blacksquare$}}
\put(922,147){\makebox(0,0){$\blacksquare$}}
\put(933,145){\makebox(0,0){$\blacksquare$}}
\put(944,144){\makebox(0,0){$\blacksquare$}}
\put(955,144){\makebox(0,0){$\blacksquare$}}
\put(966,142){\makebox(0,0){$\blacksquare$}}
\put(977,143){\makebox(0,0){$\blacksquare$}}
\put(988,141){\makebox(0,0){$\blacksquare$}}
\put(999,142){\makebox(0,0){$\blacksquare$}}
\put(1009,141){\makebox(0,0){$\blacksquare$}}
\put(1020,140){\makebox(0,0){$\blacksquare$}}
\put(1031,140){\makebox(0,0){$\blacksquare$}}
\put(1042,141){\makebox(0,0){$\blacksquare$}}
\put(1053,140){\makebox(0,0){$\blacksquare$}}
\put(1064,137){\makebox(0,0){$\blacksquare$}}
\put(191,131){\makebox(0,0){$\bullet$}}
\put(202,133){\makebox(0,0){$\bullet$}}
\put(213,159){\makebox(0,0){$\bullet$}}
\put(224,230){\makebox(0,0){$\bullet$}}
\put(235,311){\makebox(0,0){$\bullet$}}
\put(246,380){\makebox(0,0){$\bullet$}}
\put(256,424){\makebox(0,0){$\bullet$}}
\put(267,450){\makebox(0,0){$\bullet$}}
\put(278,448){\makebox(0,0){$\bullet$}}
\put(289,443){\makebox(0,0){$\bullet$}}
\put(300,437){\makebox(0,0){$\bullet$}}
\put(311,412){\makebox(0,0){$\bullet$}}
\put(322,397){\makebox(0,0){$\bullet$}}
\put(333,378){\makebox(0,0){$\bullet$}}
\put(344,351){\makebox(0,0){$\bullet$}}
\put(355,335){\makebox(0,0){$\bullet$}}
\put(366,311){\makebox(0,0){$\bullet$}}
\put(377,298){\makebox(0,0){$\bullet$}}
\put(387,281){\makebox(0,0){$\bullet$}}
\put(398,267){\makebox(0,0){$\bullet$}}
\put(409,252){\makebox(0,0){$\bullet$}}
\put(420,236){\makebox(0,0){$\bullet$}}
\put(431,234){\makebox(0,0){$\bullet$}}
\put(442,217){\makebox(0,0){$\bullet$}}
\put(453,210){\makebox(0,0){$\bullet$}}
\put(464,204){\makebox(0,0){$\bullet$}}
\put(475,201){\makebox(0,0){$\bullet$}}
\put(486,188){\makebox(0,0){$\bullet$}}
\put(497,185){\makebox(0,0){$\bullet$}}
\put(507,175){\makebox(0,0){$\bullet$}}
\put(518,173){\makebox(0,0){$\bullet$}}
\put(529,168){\makebox(0,0){$\bullet$}}
\put(540,164){\makebox(0,0){$\bullet$}}
\put(551,163){\makebox(0,0){$\bullet$}}
\put(562,157){\makebox(0,0){$\bullet$}}
\put(573,157){\makebox(0,0){$\bullet$}}
\put(584,154){\makebox(0,0){$\bullet$}}
\put(595,151){\makebox(0,0){$\bullet$}}
\put(606,151){\makebox(0,0){$\bullet$}}
\put(617,147){\makebox(0,0){$\bullet$}}
\put(628,146){\makebox(0,0){$\bullet$}}
\put(638,144){\makebox(0,0){$\bullet$}}
\put(649,143){\makebox(0,0){$\bullet$}}
\put(660,142){\makebox(0,0){$\bullet$}}
\put(671,139){\makebox(0,0){$\bullet$}}
\put(682,139){\makebox(0,0){$\bullet$}}
\put(693,138){\makebox(0,0){$\bullet$}}
\put(704,139){\makebox(0,0){$\bullet$}}
\put(715,137){\makebox(0,0){$\bullet$}}
\put(726,136){\makebox(0,0){$\bullet$}}
\put(737,136){\makebox(0,0){$\bullet$}}
\put(748,136){\makebox(0,0){$\bullet$}}
\put(758,135){\makebox(0,0){$\bullet$}}
\put(769,135){\makebox(0,0){$\bullet$}}
\put(780,135){\makebox(0,0){$\bullet$}}
\put(791,134){\makebox(0,0){$\bullet$}}
\put(802,134){\makebox(0,0){$\bullet$}}
\put(813,134){\makebox(0,0){$\bullet$}}
\put(824,132){\makebox(0,0){$\bullet$}}
\put(835,133){\makebox(0,0){$\bullet$}}
\put(846,133){\makebox(0,0){$\bullet$}}
\put(857,132){\makebox(0,0){$\bullet$}}
\put(868,132){\makebox(0,0){$\bullet$}}
\put(878,132){\makebox(0,0){$\bullet$}}
\put(889,132){\makebox(0,0){$\bullet$}}
\put(900,132){\makebox(0,0){$\bullet$}}
\put(911,131){\makebox(0,0){$\bullet$}}
\put(922,131){\makebox(0,0){$\bullet$}}
\put(933,132){\makebox(0,0){$\bullet$}}
\put(944,132){\makebox(0,0){$\bullet$}}
\put(955,132){\makebox(0,0){$\bullet$}}
\put(966,131){\makebox(0,0){$\bullet$}}
\put(977,132){\makebox(0,0){$\bullet$}}
\put(988,132){\makebox(0,0){$\bullet$}}
\put(999,131){\makebox(0,0){$\bullet$}}
\put(1009,131){\makebox(0,0){$\bullet$}}
\put(1020,131){\makebox(0,0){$\bullet$}}
\put(1031,131){\makebox(0,0){$\bullet$}}
\put(1042,131){\makebox(0,0){$\bullet$}}
\put(1053,131){\makebox(0,0){$\bullet$}}
\put(1064,131){\makebox(0,0){$\bullet$}}
\sbox{\plotpoint}{\rule[-0.400pt]{0.800pt}{0.800pt}}%
\put(191,131){\makebox(0,0){$\blacktriangle$}}
\put(202,162){\makebox(0,0){$\blacktriangle$}}
\put(213,311){\makebox(0,0){$\blacktriangle$}}
\put(224,481){\makebox(0,0){$\blacktriangle$}}
\put(235,577){\makebox(0,0){$\blacktriangle$}}
\put(246,613){\makebox(0,0){$\blacktriangle$}}
\put(256,604){\makebox(0,0){$\blacktriangle$}}
\put(267,554){\makebox(0,0){$\blacktriangle$}}
\put(278,517){\makebox(0,0){$\blacktriangle$}}
\put(289,466){\makebox(0,0){$\blacktriangle$}}
\put(300,430){\makebox(0,0){$\blacktriangle$}}
\put(311,383){\makebox(0,0){$\blacktriangle$}}
\put(322,357){\makebox(0,0){$\blacktriangle$}}
\put(333,321){\makebox(0,0){$\blacktriangle$}}
\put(344,294){\makebox(0,0){$\blacktriangle$}}
\put(355,269){\makebox(0,0){$\blacktriangle$}}
\put(366,245){\makebox(0,0){$\blacktriangle$}}
\put(377,231){\makebox(0,0){$\blacktriangle$}}
\put(387,214){\makebox(0,0){$\blacktriangle$}}
\put(398,206){\makebox(0,0){$\blacktriangle$}}
\put(409,195){\makebox(0,0){$\blacktriangle$}}
\put(420,184){\makebox(0,0){$\blacktriangle$}}
\put(431,176){\makebox(0,0){$\blacktriangle$}}
\put(442,168){\makebox(0,0){$\blacktriangle$}}
\put(453,163){\makebox(0,0){$\blacktriangle$}}
\put(464,159){\makebox(0,0){$\blacktriangle$}}
\put(475,156){\makebox(0,0){$\blacktriangle$}}
\put(486,151){\makebox(0,0){$\blacktriangle$}}
\put(497,149){\makebox(0,0){$\blacktriangle$}}
\put(507,144){\makebox(0,0){$\blacktriangle$}}
\put(518,143){\makebox(0,0){$\blacktriangle$}}
\put(529,142){\makebox(0,0){$\blacktriangle$}}
\put(540,140){\makebox(0,0){$\blacktriangle$}}
\put(551,139){\makebox(0,0){$\blacktriangle$}}
\put(562,138){\makebox(0,0){$\blacktriangle$}}
\put(573,136){\makebox(0,0){$\blacktriangle$}}
\put(584,136){\makebox(0,0){$\blacktriangle$}}
\put(595,134){\makebox(0,0){$\blacktriangle$}}
\put(606,134){\makebox(0,0){$\blacktriangle$}}
\put(617,135){\makebox(0,0){$\blacktriangle$}}
\put(628,134){\makebox(0,0){$\blacktriangle$}}
\put(638,133){\makebox(0,0){$\blacktriangle$}}
\put(649,133){\makebox(0,0){$\blacktriangle$}}
\put(660,133){\makebox(0,0){$\blacktriangle$}}
\put(671,132){\makebox(0,0){$\blacktriangle$}}
\put(682,132){\makebox(0,0){$\blacktriangle$}}
\put(693,132){\makebox(0,0){$\blacktriangle$}}
\put(704,132){\makebox(0,0){$\blacktriangle$}}
\put(715,132){\makebox(0,0){$\blacktriangle$}}
\put(726,132){\makebox(0,0){$\blacktriangle$}}
\put(737,132){\makebox(0,0){$\blacktriangle$}}
\put(748,131){\makebox(0,0){$\blacktriangle$}}
\put(758,131){\makebox(0,0){$\blacktriangle$}}
\put(769,131){\makebox(0,0){$\blacktriangle$}}
\put(780,131){\makebox(0,0){$\blacktriangle$}}
\put(791,131){\makebox(0,0){$\blacktriangle$}}
\put(802,131){\makebox(0,0){$\blacktriangle$}}
\put(813,131){\makebox(0,0){$\blacktriangle$}}
\put(824,131){\makebox(0,0){$\blacktriangle$}}
\put(835,131){\makebox(0,0){$\blacktriangle$}}
\put(846,131){\makebox(0,0){$\blacktriangle$}}
\put(857,131){\makebox(0,0){$\blacktriangle$}}
\put(868,131){\makebox(0,0){$\blacktriangle$}}
\put(878,131){\makebox(0,0){$\blacktriangle$}}
\put(889,131){\makebox(0,0){$\blacktriangle$}}
\put(900,131){\makebox(0,0){$\blacktriangle$}}
\put(911,131){\makebox(0,0){$\blacktriangle$}}
\put(922,131){\makebox(0,0){$\blacktriangle$}}
\put(933,131){\makebox(0,0){$\blacktriangle$}}
\put(944,131){\makebox(0,0){$\blacktriangle$}}
\put(955,131){\makebox(0,0){$\blacktriangle$}}
\put(966,131){\makebox(0,0){$\blacktriangle$}}
\put(977,131){\makebox(0,0){$\blacktriangle$}}
\put(988,131){\makebox(0,0){$\blacktriangle$}}
\put(999,131){\makebox(0,0){$\blacktriangle$}}
\put(1009,131){\makebox(0,0){$\blacktriangle$}}
\put(1020,131){\makebox(0,0){$\blacktriangle$}}
\put(1031,131){\makebox(0,0){$\blacktriangle$}}
\put(1042,131){\makebox(0,0){$\blacktriangle$}}
\put(1053,131){\makebox(0,0){$\blacktriangle$}}
\put(1064,131){\makebox(0,0){$\blacktriangle$}}
\sbox{\plotpoint}{\rule[-0.500pt]{1.000pt}{1.000pt}}%
\put(191,133){\makebox(0,0){$\blacktriangledown$}}
\put(202,255){\makebox(0,0){$\blacktriangledown$}}
\put(213,563){\makebox(0,0){$\blacktriangledown$}}
\put(224,749){\makebox(0,0){$\blacktriangledown$}}
\put(235,762){\makebox(0,0){$\blacktriangledown$}}
\put(246,715){\makebox(0,0){$\blacktriangledown$}}
\put(256,630){\makebox(0,0){$\blacktriangledown$}}
\put(267,559){\makebox(0,0){$\blacktriangledown$}}
\put(278,474){\makebox(0,0){$\blacktriangledown$}}
\put(289,415){\makebox(0,0){$\blacktriangledown$}}
\put(300,365){\makebox(0,0){$\blacktriangledown$}}
\put(311,324){\makebox(0,0){$\blacktriangledown$}}
\put(322,281){\makebox(0,0){$\blacktriangledown$}}
\put(333,258){\makebox(0,0){$\blacktriangledown$}}
\put(344,227){\makebox(0,0){$\blacktriangledown$}}
\put(355,214){\makebox(0,0){$\blacktriangledown$}}
\put(366,194){\makebox(0,0){$\blacktriangledown$}}
\put(377,184){\makebox(0,0){$\blacktriangledown$}}
\put(387,172){\makebox(0,0){$\blacktriangledown$}}
\put(398,167){\makebox(0,0){$\blacktriangledown$}}
\put(409,159){\makebox(0,0){$\blacktriangledown$}}
\put(420,154){\makebox(0,0){$\blacktriangledown$}}
\put(431,149){\makebox(0,0){$\blacktriangledown$}}
\put(442,147){\makebox(0,0){$\blacktriangledown$}}
\put(453,142){\makebox(0,0){$\blacktriangledown$}}
\put(464,141){\makebox(0,0){$\blacktriangledown$}}
\put(475,139){\makebox(0,0){$\blacktriangledown$}}
\put(486,137){\makebox(0,0){$\blacktriangledown$}}
\put(497,137){\makebox(0,0){$\blacktriangledown$}}
\put(507,134){\makebox(0,0){$\blacktriangledown$}}
\put(518,134){\makebox(0,0){$\blacktriangledown$}}
\put(529,133){\makebox(0,0){$\blacktriangledown$}}
\put(540,133){\makebox(0,0){$\blacktriangledown$}}
\put(551,132){\makebox(0,0){$\blacktriangledown$}}
\put(562,132){\makebox(0,0){$\blacktriangledown$}}
\put(573,132){\makebox(0,0){$\blacktriangledown$}}
\put(584,132){\makebox(0,0){$\blacktriangledown$}}
\put(595,132){\makebox(0,0){$\blacktriangledown$}}
\put(606,132){\makebox(0,0){$\blacktriangledown$}}
\put(617,132){\makebox(0,0){$\blacktriangledown$}}
\put(628,131){\makebox(0,0){$\blacktriangledown$}}
\put(638,131){\makebox(0,0){$\blacktriangledown$}}
\put(649,131){\makebox(0,0){$\blacktriangledown$}}
\put(660,131){\makebox(0,0){$\blacktriangledown$}}
\put(671,131){\makebox(0,0){$\blacktriangledown$}}
\put(682,131){\makebox(0,0){$\blacktriangledown$}}
\put(693,131){\makebox(0,0){$\blacktriangledown$}}
\put(704,131){\makebox(0,0){$\blacktriangledown$}}
\put(715,131){\makebox(0,0){$\blacktriangledown$}}
\put(726,131){\makebox(0,0){$\blacktriangledown$}}
\put(737,131){\makebox(0,0){$\blacktriangledown$}}
\put(748,131){\makebox(0,0){$\blacktriangledown$}}
\put(758,131){\makebox(0,0){$\blacktriangledown$}}
\put(769,131){\makebox(0,0){$\blacktriangledown$}}
\put(780,131){\makebox(0,0){$\blacktriangledown$}}
\put(791,131){\makebox(0,0){$\blacktriangledown$}}
\put(802,131){\makebox(0,0){$\blacktriangledown$}}
\put(813,131){\makebox(0,0){$\blacktriangledown$}}
\put(824,131){\makebox(0,0){$\blacktriangledown$}}
\put(835,131){\makebox(0,0){$\blacktriangledown$}}
\put(846,131){\makebox(0,0){$\blacktriangledown$}}
\put(857,131){\makebox(0,0){$\blacktriangledown$}}
\put(868,131){\makebox(0,0){$\blacktriangledown$}}
\put(878,131){\makebox(0,0){$\blacktriangledown$}}
\put(889,131){\makebox(0,0){$\blacktriangledown$}}
\put(900,131){\makebox(0,0){$\blacktriangledown$}}
\put(911,131){\makebox(0,0){$\blacktriangledown$}}
\put(922,131){\makebox(0,0){$\blacktriangledown$}}
\put(933,131){\makebox(0,0){$\blacktriangledown$}}
\put(944,131){\makebox(0,0){$\blacktriangledown$}}
\put(955,131){\makebox(0,0){$\blacktriangledown$}}
\put(966,131){\makebox(0,0){$\blacktriangledown$}}
\put(977,131){\makebox(0,0){$\blacktriangledown$}}
\put(988,131){\makebox(0,0){$\blacktriangledown$}}
\put(999,131){\makebox(0,0){$\blacktriangledown$}}
\put(1009,131){\makebox(0,0){$\blacktriangledown$}}
\put(1020,131){\makebox(0,0){$\blacktriangledown$}}
\put(1031,131){\makebox(0,0){$\blacktriangledown$}}
\put(1042,131){\makebox(0,0){$\blacktriangledown$}}
\put(1053,131){\makebox(0,0){$\blacktriangledown$}}
\put(1064,131){\makebox(0,0){$\blacktriangledown$}}
\sbox{\plotpoint}{\rule[-0.200pt]{0.400pt}{0.400pt}}%
\put(191.0,131.0){\rule[-0.200pt]{0.400pt}{175.375pt}}
\put(191.0,131.0){\rule[-0.200pt]{210.306pt}{0.400pt}}
\put(1064.0,131.0){\rule[-0.200pt]{0.400pt}{175.375pt}}
\put(191.0,859.0){\rule[-0.200pt]{210.306pt}{0.400pt}}
\end{picture}

\begin{picture}(1125,900)(0,0)
\put(191.0,131.0){\rule[-0.200pt]{4.818pt}{0.400pt}}
\put(171,131){\makebox(0,0)[r]{ 0}}
\put(1044.0,131.0){\rule[-0.200pt]{4.818pt}{0.400pt}}
\put(191.0,222.0){\rule[-0.200pt]{4.818pt}{0.400pt}}
\put(171,222){\makebox(0,0)[r]{ 0.02}}
\put(1044.0,222.0){\rule[-0.200pt]{4.818pt}{0.400pt}}
\put(191.0,313.0){\rule[-0.200pt]{4.818pt}{0.400pt}}
\put(171,313){\makebox(0,0)[r]{ 0.04}}
\put(1044.0,313.0){\rule[-0.200pt]{4.818pt}{0.400pt}}
\put(191.0,404.0){\rule[-0.200pt]{4.818pt}{0.400pt}}
\put(171,404){\makebox(0,0)[r]{ 0.06}}
\put(1044.0,404.0){\rule[-0.200pt]{4.818pt}{0.400pt}}
\put(191.0,495.0){\rule[-0.200pt]{4.818pt}{0.400pt}}
\put(171,495){\makebox(0,0)[r]{ 0.08}}
\put(1044.0,495.0){\rule[-0.200pt]{4.818pt}{0.400pt}}
\put(191.0,586.0){\rule[-0.200pt]{4.818pt}{0.400pt}}
\put(171,586){\makebox(0,0)[r]{ 0.1}}
\put(1044.0,586.0){\rule[-0.200pt]{4.818pt}{0.400pt}}
\put(191.0,677.0){\rule[-0.200pt]{4.818pt}{0.400pt}}
\put(171,677){\makebox(0,0)[r]{ 0.12}}
\put(1044.0,677.0){\rule[-0.200pt]{4.818pt}{0.400pt}}
\put(191.0,768.0){\rule[-0.200pt]{4.818pt}{0.400pt}}
\put(171,768){\makebox(0,0)[r]{ 0.14}}
\put(1044.0,768.0){\rule[-0.200pt]{4.818pt}{0.400pt}}
\put(191.0,859.0){\rule[-0.200pt]{4.818pt}{0.400pt}}
\put(171,859){\makebox(0,0)[r]{ 0.16}}
\put(1044.0,859.0){\rule[-0.200pt]{4.818pt}{0.400pt}}
\put(191.0,131.0){\rule[-0.200pt]{0.400pt}{4.818pt}}
\put(191,90){\makebox(0,0){ 0}}
\put(191.0,839.0){\rule[-0.200pt]{0.400pt}{4.818pt}}
\put(409.0,131.0){\rule[-0.200pt]{0.400pt}{4.818pt}}
\put(409,90){\makebox(0,0){ 1000}}
\put(409.0,839.0){\rule[-0.200pt]{0.400pt}{4.818pt}}
\put(628.0,131.0){\rule[-0.200pt]{0.400pt}{4.818pt}}
\put(628,90){\makebox(0,0){ 2000}}
\put(628.0,839.0){\rule[-0.200pt]{0.400pt}{4.818pt}}
\put(846.0,131.0){\rule[-0.200pt]{0.400pt}{4.818pt}}
\put(846,90){\makebox(0,0){ 3000}}
\put(846.0,839.0){\rule[-0.200pt]{0.400pt}{4.818pt}}
\put(1064.0,131.0){\rule[-0.200pt]{0.400pt}{4.818pt}}
\put(1064,90){\makebox(0,0){ 4000}}
\put(1064.0,839.0){\rule[-0.200pt]{0.400pt}{4.818pt}}
\put(191.0,131.0){\rule[-0.200pt]{0.400pt}{175.375pt}}
\put(191.0,131.0){\rule[-0.200pt]{210.306pt}{0.400pt}}
\put(1064.0,131.0){\rule[-0.200pt]{0.400pt}{175.375pt}}
\put(191.0,859.0){\rule[-0.200pt]{210.306pt}{0.400pt}}
\put(20,495){\makebox(0,0){${P(t_1)}p^{\gamma}$}}
\put(627,29){\makebox(0,0){${t_1}p^{\delta}$}}
\put(628,677){\makebox(0,0)[l]{$\gamma=-1$}}
\put(628,586){\makebox(0,0)[l]{$\delta=1$}}
\put(955,768){\makebox(0,0)[l]{(b)}}
\put(191,131){\makebox(0,0){$\blacksquare$}}
\put(198,131){\makebox(0,0){$\blacksquare$}}
\put(204,131){\makebox(0,0){$\blacksquare$}}
\put(211,142){\makebox(0,0){$\blacksquare$}}
\put(217,168){\makebox(0,0){$\blacksquare$}}
\put(224,225){\makebox(0,0){$\blacksquare$}}
\put(230,304){\makebox(0,0){$\blacksquare$}}
\put(237,397){\makebox(0,0){$\blacksquare$}}
\put(243,480){\makebox(0,0){$\blacksquare$}}
\put(250,574){\makebox(0,0){$\blacksquare$}}
\put(256,645){\makebox(0,0){$\blacksquare$}}
\put(263,678){\makebox(0,0){$\blacksquare$}}
\put(270,768){\makebox(0,0){$\blacksquare$}}
\put(276,791){\makebox(0,0){$\blacksquare$}}
\put(283,806){\makebox(0,0){$\blacksquare$}}
\put(289,829){\makebox(0,0){$\blacksquare$}}
\put(296,836){\makebox(0,0){$\blacksquare$}}
\put(302,841){\makebox(0,0){$\blacksquare$}}
\put(309,817){\makebox(0,0){$\blacksquare$}}
\put(315,817){\makebox(0,0){$\blacksquare$}}
\put(322,798){\makebox(0,0){$\blacksquare$}}
\put(328,783){\makebox(0,0){$\blacksquare$}}
\put(335,754){\makebox(0,0){$\blacksquare$}}
\put(342,764){\makebox(0,0){$\blacksquare$}}
\put(348,700){\makebox(0,0){$\blacksquare$}}
\put(355,693){\makebox(0,0){$\blacksquare$}}
\put(361,675){\makebox(0,0){$\blacksquare$}}
\put(368,658){\makebox(0,0){$\blacksquare$}}
\put(374,629){\makebox(0,0){$\blacksquare$}}
\put(381,589){\makebox(0,0){$\blacksquare$}}
\put(387,581){\makebox(0,0){$\blacksquare$}}
\put(394,566){\makebox(0,0){$\blacksquare$}}
\put(401,548){\makebox(0,0){$\blacksquare$}}
\put(407,532){\makebox(0,0){$\blacksquare$}}
\put(414,499){\makebox(0,0){$\blacksquare$}}
\put(420,478){\makebox(0,0){$\blacksquare$}}
\put(427,476){\makebox(0,0){$\blacksquare$}}
\put(433,443){\makebox(0,0){$\blacksquare$}}
\put(440,446){\makebox(0,0){$\blacksquare$}}
\put(446,412){\makebox(0,0){$\blacksquare$}}
\put(453,397){\makebox(0,0){$\blacksquare$}}
\put(459,388){\makebox(0,0){$\blacksquare$}}
\put(466,370){\makebox(0,0){$\blacksquare$}}
\put(473,365){\makebox(0,0){$\blacksquare$}}
\put(479,350){\makebox(0,0){$\blacksquare$}}
\put(486,354){\makebox(0,0){$\blacksquare$}}
\put(492,330){\makebox(0,0){$\blacksquare$}}
\put(499,325){\makebox(0,0){$\blacksquare$}}
\put(505,315){\makebox(0,0){$\blacksquare$}}
\put(512,308){\makebox(0,0){$\blacksquare$}}
\put(518,288){\makebox(0,0){$\blacksquare$}}
\put(525,286){\makebox(0,0){$\blacksquare$}}
\put(531,277){\makebox(0,0){$\blacksquare$}}
\put(538,264){\makebox(0,0){$\blacksquare$}}
\put(545,257){\makebox(0,0){$\blacksquare$}}
\put(551,253){\makebox(0,0){$\blacksquare$}}
\put(558,245){\makebox(0,0){$\blacksquare$}}
\put(564,237){\makebox(0,0){$\blacksquare$}}
\put(571,236){\makebox(0,0){$\blacksquare$}}
\put(577,235){\makebox(0,0){$\blacksquare$}}
\put(584,225){\makebox(0,0){$\blacksquare$}}
\put(590,222){\makebox(0,0){$\blacksquare$}}
\put(597,211){\makebox(0,0){$\blacksquare$}}
\put(603,212){\makebox(0,0){$\blacksquare$}}
\put(610,212){\makebox(0,0){$\blacksquare$}}
\put(617,205){\makebox(0,0){$\blacksquare$}}
\put(623,196){\makebox(0,0){$\blacksquare$}}
\put(630,203){\makebox(0,0){$\blacksquare$}}
\put(636,193){\makebox(0,0){$\blacksquare$}}
\put(643,187){\makebox(0,0){$\blacksquare$}}
\put(649,186){\makebox(0,0){$\blacksquare$}}
\put(656,181){\makebox(0,0){$\blacksquare$}}
\put(662,185){\makebox(0,0){$\blacksquare$}}
\put(669,174){\makebox(0,0){$\blacksquare$}}
\put(676,178){\makebox(0,0){$\blacksquare$}}
\put(682,177){\makebox(0,0){$\blacksquare$}}
\put(689,169){\makebox(0,0){$\blacksquare$}}
\put(695,171){\makebox(0,0){$\blacksquare$}}
\put(702,173){\makebox(0,0){$\blacksquare$}}
\put(708,169){\makebox(0,0){$\blacksquare$}}
\put(715,159){\makebox(0,0){$\blacksquare$}}
\put(721,159){\makebox(0,0){$\blacksquare$}}
\put(728,161){\makebox(0,0){$\blacksquare$}}
\put(734,156){\makebox(0,0){$\blacksquare$}}
\put(741,160){\makebox(0,0){$\blacksquare$}}
\put(748,157){\makebox(0,0){$\blacksquare$}}
\put(754,160){\makebox(0,0){$\blacksquare$}}
\put(761,154){\makebox(0,0){$\blacksquare$}}
\put(767,149){\makebox(0,0){$\blacksquare$}}
\put(774,151){\makebox(0,0){$\blacksquare$}}
\put(780,152){\makebox(0,0){$\blacksquare$}}
\put(787,151){\makebox(0,0){$\blacksquare$}}
\put(793,150){\makebox(0,0){$\blacksquare$}}
\put(800,149){\makebox(0,0){$\blacksquare$}}
\put(806,145){\makebox(0,0){$\blacksquare$}}
\put(813,146){\makebox(0,0){$\blacksquare$}}
\put(820,145){\makebox(0,0){$\blacksquare$}}
\put(826,144){\makebox(0,0){$\blacksquare$}}
\put(833,146){\makebox(0,0){$\blacksquare$}}
\put(839,141){\makebox(0,0){$\blacksquare$}}
\put(846,142){\makebox(0,0){$\blacksquare$}}
\put(852,139){\makebox(0,0){$\blacksquare$}}
\put(859,141){\makebox(0,0){$\blacksquare$}}
\put(865,139){\makebox(0,0){$\blacksquare$}}
\put(872,141){\makebox(0,0){$\blacksquare$}}
\put(878,140){\makebox(0,0){$\blacksquare$}}
\put(885,138){\makebox(0,0){$\blacksquare$}}
\put(892,136){\makebox(0,0){$\blacksquare$}}
\put(898,138){\makebox(0,0){$\blacksquare$}}
\put(905,136){\makebox(0,0){$\blacksquare$}}
\put(911,139){\makebox(0,0){$\blacksquare$}}
\put(918,135){\makebox(0,0){$\blacksquare$}}
\put(924,135){\makebox(0,0){$\blacksquare$}}
\put(931,136){\makebox(0,0){$\blacksquare$}}
\put(937,136){\makebox(0,0){$\blacksquare$}}
\put(944,136){\makebox(0,0){$\blacksquare$}}
\put(951,136){\makebox(0,0){$\blacksquare$}}
\put(957,136){\makebox(0,0){$\blacksquare$}}
\put(964,136){\makebox(0,0){$\blacksquare$}}
\put(970,135){\makebox(0,0){$\blacksquare$}}
\put(977,135){\makebox(0,0){$\blacksquare$}}
\put(983,134){\makebox(0,0){$\blacksquare$}}
\put(990,135){\makebox(0,0){$\blacksquare$}}
\put(996,134){\makebox(0,0){$\blacksquare$}}
\put(1003,135){\makebox(0,0){$\blacksquare$}}
\put(1009,134){\makebox(0,0){$\blacksquare$}}
\put(1016,135){\makebox(0,0){$\blacksquare$}}
\put(1023,134){\makebox(0,0){$\blacksquare$}}
\put(1029,133){\makebox(0,0){$\blacksquare$}}
\put(1036,134){\makebox(0,0){$\blacksquare$}}
\put(1042,134){\makebox(0,0){$\blacksquare$}}
\put(1049,133){\makebox(0,0){$\blacksquare$}}
\put(1055,133){\makebox(0,0){$\blacksquare$}}
\put(1062,133){\makebox(0,0){$\blacksquare$}}
\put(191,131){\makebox(0,0){$\bullet$}}
\put(204,134){\makebox(0,0){$\bullet$}}
\put(217,191){\makebox(0,0){$\bullet$}}
\put(230,348){\makebox(0,0){$\bullet$}}
\put(243,524){\makebox(0,0){$\bullet$}}
\put(256,675){\makebox(0,0){$\bullet$}}
\put(270,772){\makebox(0,0){$\bullet$}}
\put(283,829){\makebox(0,0){$\bullet$}}
\put(296,825){\makebox(0,0){$\bullet$}}
\put(309,814){\makebox(0,0){$\bullet$}}
\put(322,800){\makebox(0,0){$\bullet$}}
\put(335,746){\makebox(0,0){$\bullet$}}
\put(348,713){\makebox(0,0){$\bullet$}}
\put(361,672){\makebox(0,0){$\bullet$}}
\put(374,612){\makebox(0,0){$\bullet$}}
\put(387,576){\makebox(0,0){$\bullet$}}
\put(401,525){\makebox(0,0){$\bullet$}}
\put(414,497){\makebox(0,0){$\bullet$}}
\put(427,459){\makebox(0,0){$\bullet$}}
\put(440,428){\makebox(0,0){$\bullet$}}
\put(453,397){\makebox(0,0){$\bullet$}}
\put(466,361){\makebox(0,0){$\bullet$}}
\put(479,357){\makebox(0,0){$\bullet$}}
\put(492,320){\makebox(0,0){$\bullet$}}
\put(505,303){\makebox(0,0){$\bullet$}}
\put(518,291){\makebox(0,0){$\bullet$}}
\put(531,284){\makebox(0,0){$\bullet$}}
\put(545,256){\makebox(0,0){$\bullet$}}
\put(558,248){\makebox(0,0){$\bullet$}}
\put(571,228){\makebox(0,0){$\bullet$}}
\put(584,222){\makebox(0,0){$\bullet$}}
\put(597,211){\makebox(0,0){$\bullet$}}
\put(610,203){\makebox(0,0){$\bullet$}}
\put(623,201){\makebox(0,0){$\bullet$}}
\put(636,189){\makebox(0,0){$\bullet$}}
\put(649,187){\makebox(0,0){$\bullet$}}
\put(662,180){\makebox(0,0){$\bullet$}}
\put(676,176){\makebox(0,0){$\bullet$}}
\put(689,176){\makebox(0,0){$\bullet$}}
\put(702,166){\makebox(0,0){$\bullet$}}
\put(715,164){\makebox(0,0){$\bullet$}}
\put(728,159){\makebox(0,0){$\bullet$}}
\put(741,158){\makebox(0,0){$\bullet$}}
\put(754,155){\makebox(0,0){$\bullet$}}
\put(767,149){\makebox(0,0){$\bullet$}}
\put(780,148){\makebox(0,0){$\bullet$}}
\put(793,146){\makebox(0,0){$\bullet$}}
\put(806,148){\makebox(0,0){$\bullet$}}
\put(820,145){\makebox(0,0){$\bullet$}}
\put(833,142){\makebox(0,0){$\bullet$}}
\put(846,141){\makebox(0,0){$\bullet$}}
\put(859,141){\makebox(0,0){$\bullet$}}
\put(872,139){\makebox(0,0){$\bullet$}}
\put(885,140){\makebox(0,0){$\bullet$}}
\put(898,140){\makebox(0,0){$\bullet$}}
\put(911,139){\makebox(0,0){$\bullet$}}
\put(924,138){\makebox(0,0){$\bullet$}}
\put(937,137){\makebox(0,0){$\bullet$}}
\put(951,134){\makebox(0,0){$\bullet$}}
\put(964,136){\makebox(0,0){$\bullet$}}
\put(977,135){\makebox(0,0){$\bullet$}}
\put(990,134){\makebox(0,0){$\bullet$}}
\put(1003,134){\makebox(0,0){$\bullet$}}
\put(1016,133){\makebox(0,0){$\bullet$}}
\put(1029,133){\makebox(0,0){$\bullet$}}
\put(1042,134){\makebox(0,0){$\bullet$}}
\put(1055,132){\makebox(0,0){$\bullet$}}
\sbox{\plotpoint}{\rule[-0.400pt]{0.800pt}{0.800pt}}%
\put(191,131){\makebox(0,0){$\blacktriangle$}}
\put(211,176){\makebox(0,0){$\blacktriangle$}}
\put(230,394){\makebox(0,0){$\blacktriangle$}}
\put(250,642){\makebox(0,0){$\blacktriangle$}}
\put(270,781){\makebox(0,0){$\blacktriangle$}}
\put(289,834){\makebox(0,0){$\blacktriangle$}}
\put(309,820){\makebox(0,0){$\blacktriangle$}}
\put(328,748){\makebox(0,0){$\blacktriangle$}}
\put(348,694){\makebox(0,0){$\blacktriangle$}}
\put(368,619){\makebox(0,0){$\blacktriangle$}}
\put(387,567){\makebox(0,0){$\blacktriangle$}}
\put(407,499){\makebox(0,0){$\blacktriangle$}}
\put(427,460){\makebox(0,0){$\blacktriangle$}}
\put(446,408){\makebox(0,0){$\blacktriangle$}}
\put(466,369){\makebox(0,0){$\blacktriangle$}}
\put(486,332){\makebox(0,0){$\blacktriangle$}}
\put(505,297){\makebox(0,0){$\blacktriangle$}}
\put(525,277){\makebox(0,0){$\blacktriangle$}}
\put(545,251){\makebox(0,0){$\blacktriangle$}}
\put(564,241){\makebox(0,0){$\blacktriangle$}}
\put(584,225){\makebox(0,0){$\blacktriangle$}}
\put(603,208){\makebox(0,0){$\blacktriangle$}}
\put(623,197){\makebox(0,0){$\blacktriangle$}}
\put(643,185){\makebox(0,0){$\blacktriangle$}}
\put(662,177){\makebox(0,0){$\blacktriangle$}}
\put(682,171){\makebox(0,0){$\blacktriangle$}}
\put(702,167){\makebox(0,0){$\blacktriangle$}}
\put(721,160){\makebox(0,0){$\blacktriangle$}}
\put(741,157){\makebox(0,0){$\blacktriangle$}}
\put(761,150){\makebox(0,0){$\blacktriangle$}}
\put(780,149){\makebox(0,0){$\blacktriangle$}}
\put(800,148){\makebox(0,0){$\blacktriangle$}}
\put(820,144){\makebox(0,0){$\blacktriangle$}}
\put(839,143){\makebox(0,0){$\blacktriangle$}}
\put(859,141){\makebox(0,0){$\blacktriangle$}}
\put(878,139){\makebox(0,0){$\blacktriangle$}}
\put(898,138){\makebox(0,0){$\blacktriangle$}}
\put(918,136){\makebox(0,0){$\blacktriangle$}}
\put(937,135){\makebox(0,0){$\blacktriangle$}}
\put(957,136){\makebox(0,0){$\blacktriangle$}}
\put(977,135){\makebox(0,0){$\blacktriangle$}}
\put(996,133){\makebox(0,0){$\blacktriangle$}}
\put(1016,134){\makebox(0,0){$\blacktriangle$}}
\put(1036,134){\makebox(0,0){$\blacktriangle$}}
\put(1055,133){\makebox(0,0){$\blacktriangle$}}
\sbox{\plotpoint}{\rule[-0.500pt]{1.000pt}{1.000pt}}%
\put(191,133){\makebox(0,0){$\blacktriangledown$}}
\put(217,267){\makebox(0,0){$\blacktriangledown$}}
\put(243,603){\makebox(0,0){$\blacktriangledown$}}
\put(270,807){\makebox(0,0){$\blacktriangledown$}}
\put(296,821){\makebox(0,0){$\blacktriangledown$}}
\put(322,770){\makebox(0,0){$\blacktriangledown$}}
\put(348,676){\makebox(0,0){$\blacktriangledown$}}
\put(374,599){\makebox(0,0){$\blacktriangledown$}}
\put(401,506){\makebox(0,0){$\blacktriangledown$}}
\put(427,441){\makebox(0,0){$\blacktriangledown$}}
\put(453,386){\makebox(0,0){$\blacktriangledown$}}
\put(479,342){\makebox(0,0){$\blacktriangledown$}}
\put(505,295){\makebox(0,0){$\blacktriangledown$}}
\put(531,269){\makebox(0,0){$\blacktriangledown$}}
\put(558,236){\makebox(0,0){$\blacktriangledown$}}
\put(584,222){\makebox(0,0){$\blacktriangledown$}}
\put(610,200){\makebox(0,0){$\blacktriangledown$}}
\put(636,189){\makebox(0,0){$\blacktriangledown$}}
\put(662,176){\makebox(0,0){$\blacktriangledown$}}
\put(689,170){\makebox(0,0){$\blacktriangledown$}}
\put(715,162){\makebox(0,0){$\blacktriangledown$}}
\put(741,156){\makebox(0,0){$\blacktriangledown$}}
\put(767,150){\makebox(0,0){$\blacktriangledown$}}
\put(793,148){\makebox(0,0){$\blacktriangledown$}}
\put(820,143){\makebox(0,0){$\blacktriangledown$}}
\put(846,142){\makebox(0,0){$\blacktriangledown$}}
\put(872,139){\makebox(0,0){$\blacktriangledown$}}
\put(898,138){\makebox(0,0){$\blacktriangledown$}}
\put(924,137){\makebox(0,0){$\blacktriangledown$}}
\put(951,135){\makebox(0,0){$\blacktriangledown$}}
\put(977,134){\makebox(0,0){$\blacktriangledown$}}
\put(1003,134){\makebox(0,0){$\blacktriangledown$}}
\put(1029,134){\makebox(0,0){$\blacktriangledown$}}
\put(1055,132){\makebox(0,0){$\blacktriangledown$}}
\sbox{\plotpoint}{\rule[-0.200pt]{0.400pt}{0.400pt}}%
\put(191.0,131.0){\rule[-0.200pt]{0.400pt}{175.375pt}}
\put(191.0,131.0){\rule[-0.200pt]{210.306pt}{0.400pt}}
\put(1064.0,131.0){\rule[-0.200pt]{0.400pt}{175.375pt}}
\put(191.0,859.0){\rule[-0.200pt]{210.306pt}{0.400pt}}
\end{picture}

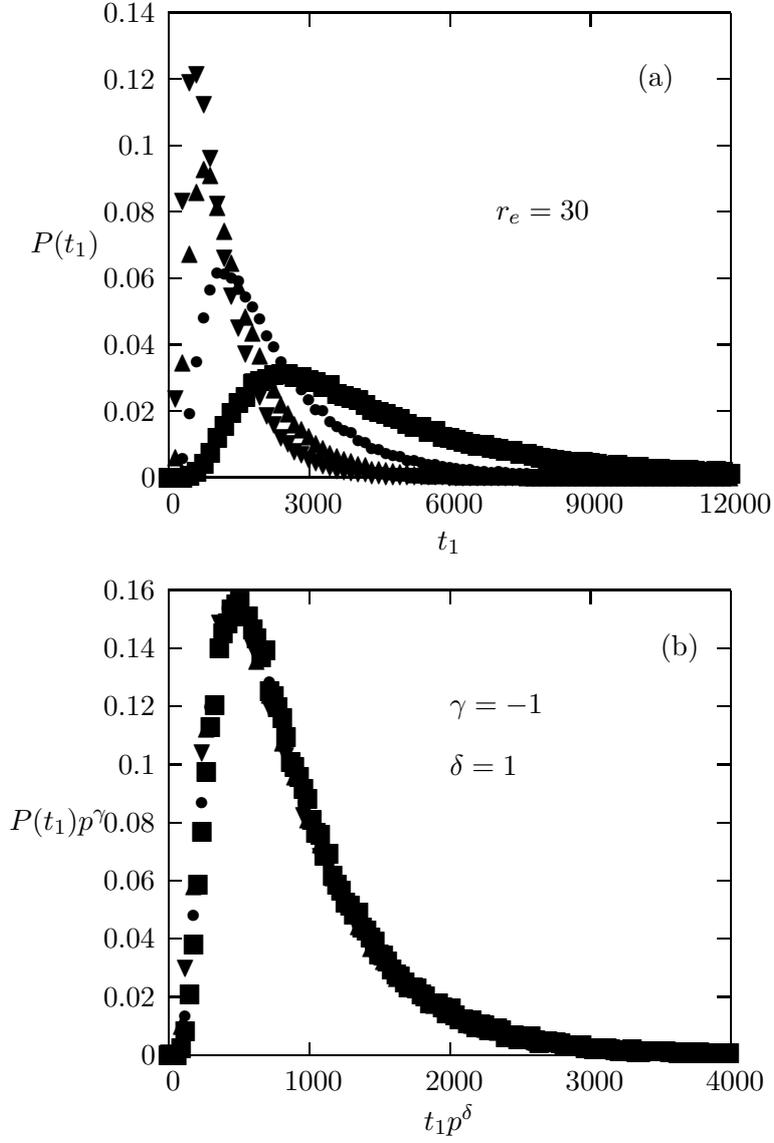
\noindent {\bf Fig-4.} The distribution (a) of first passage time ($t_1$) and
its scaling (b). Different symbol correspond to different values of jump
probability ($p$) in three dimensions. $p=0.2$($\blacksquare$), $p=0.4(\bullet)$, 
$p=0.6(\blacktriangle$) and $p=0.8(\blacktriangledown)$.
Here, $N_s = 10^5$, $N_t=10^5$ and $r_e=30$.

\end{document}